\documentclass[12pt,a4paper,aps]{revtex4}

\usepackage{epsfig}
\usepackage{graphicx}
\usepackage{amsmath,amssymb}
\usepackage{color}
\usepackage[utf8]{inputenc}

\parskip=\medskipamount

\newcommand{\eq}[1]{(\ref{#1})}
\newcommand{\fig}[1]{Fig. \ref{#1}}

\newcommand{\be}{\begin{equation}}
\newcommand{\ee}{\end{equation}}

\newcommand\disp{\displaystyle}

\newcommand{\appropto}{\mathrel{\vcenter{
  \offinterlineskip\halign{\hfil$##$\cr
    \propto\cr\noalign{\kern2pt}\sim\cr\noalign{\kern-2pt}}}}}

\begin{document}

\title{Fixman problem revisited: When fluctuations of inflated ideal polymer loop are non-Gaussian?}

\author{Sergei Nechaev}

\affiliation{Interdisciplinary Scientific Center Poncelet (CNRS IRL 2615), 119002 Moscow, Russia \\ P.N. Lebedev Physical Institute RAS, 119991 Moscow, Russia}

\author{Alexander Valov}

\affiliation{N.N. Semenov Institute of Chemical Physics RAS, 119991 Moscow, Russia}

\begin{abstract}

We consider statistics of a planar ideal polymer loop of length $L$ in a large deviation regime, when a  gyration radius, $R_g$, is slightly less than the radius of a fully inflated ring, $\frac{L}{2\pi}$. Specifically, we study analytically and via off-lattice Monte-Carlo simulations relative fluctuations of chain monomers in ensemble of Brownian loops. We have shown that these fluctuations in the regime with fixed large gyration radius are Gaussian with the critical exponent $\gamma = \frac{1}{2}$. However, if we insert inside the inflated loop the impenetrable disc of radius $R_d=R_g$, the fluctuations become non-Gaussian with the critical exponent $\gamma=\frac{1}{3}$ typical for the Kardar-Parisi-Zhang universality class.

\end{abstract}

\maketitle

\section{Introduction}

The classical problem in statistics of ideal polymers, formulated and solved by M. Fixmann in his seminal paper ``Radius of Gyration of Polymer Chain" ''\cite{fixm}, deals with the computation of the partition function, $Z_N(R_g)$, of $N$-step ideal random walk with the gyration radius, $R_g$. Later, the same problem has been reconsidered by many researchers using variety of approaches -- see, for example \cite{fix-f1,fix-f2}. In all models which do not account for volume interactions, the free energy, $F(R_g,N)$, has the following asymptotic form
\be
F(R_g,N) = \begin{cases} \disp c_1\frac{R_g^2}{Na^2} & \mbox{for $R_g^2 \sim Na^2$} \medskip \\ \disp c_2\frac{Na^2}{R_g^2} & \mbox{for $R_g^2\ll Na^2$} \end{cases}
\label{1}
\ee
where $c_1, c_2$ are model-dependent numerical constants and $a$ is the monomer size. The behavior \eq{1} is qualitatively clear: in the ``non-compressed'' regime, when the random walk is nearly free with the standard Brownian motion scaling, $R_g^2 \sim Na^2$, the distribution of the gyration radius is Gaussian; while in the ``strongly compressed'' regime, $R_g^2 \ll Na^2$, one can regard a polymer loop as random walk confined in a bounding box of typical size $R_g$. In that regime the free energy, $F$, can be estimated as $F\sim N\lambda_{min}$, where $\lambda_{min}$ is the smallest (in the absolute value) eigenvalue, $\lambda_{min} \sim\frac{a^2}{R_g^2}$, of the corresponding diffusion equation.

Despite the relation \eq{1} describes main asymptotic regimes of a polymer loop with a fixed gyration radius, $R_g$, on may wonder what happens to the closed ideal polymer chain in a strongly inflated regime, when $R_g \lesssim \frac{Na}{2\pi}$, i.e. a polymer chain is nearly a perfect ring. Extending the computations of M. Fixman to that case, one can easily check that fluctuations of the gyration radius are still Gaussian for $N\gg 1$. More interesting question concerns the local fluctuational behavior of individual monomers of a strongly inflated ideal polymer loop. As we shall see, depending on imposed boundary conditions, one can detect different scaling regimes.

We consider the 2D polymer problem, for which we compute the distribution function, $Z_N(r|R_g)$, of a particular monomer located at the point ${\bf r}$ of an ideal polymer ring in a plane with a fixed gyration radius, $R_g$. We pay attention to a specific limit of inflated loops, when the gyration radius, $R_g$, scales linearly with the chain length, $N$, i.e. $R_g = c Na$ (definitely, $c<\frac{1}{2\pi a}$).

The goal of our consideration is to highlight the simultaneous role of path stretching and imposed geometric constraints (boundary conditions). We evaluate the partition function in the limit $N\gg 1$ for two models: (i) the strongly inflated polymer loop without any boundary conditions, and (ii) the  polymer loop ``leaning'' on an large impenetrable disc of radius $R_d\le R_g$ placed inside a polymer ring. In the model (i) the partition function can be obtained by the summation of all fluctuational modes of the inflated ideal loop, and has the standard Gaussian distribution, while in the model (ii) the imposed boundary constraints prohibit the long-range fluctuations of the loop, which manifest themselves in emergence of the non-Gaussian fluctuational regime. We demonstrate that the fluctuations of the inflated ideal polymer loop supported from inside by an impenetrable disc, are controlled by the Kardar-Parisi-Zhang (KPZ) exponent $\gamma=\frac{1}{3}$. Shrinking $R_d$, and keeping the length of the polymer unchanged, we restore the Gaussian fluctuations of chain monomers with the critical exponent $\gamma=\frac{1}{2}$.

It is noteworthy that, as it has been shown in \cite{gorsky}, the KPZ scaling for fluctuations in the ensemble of inflated (or ``stretched'') random loops evading the disc, is the key property that ensures the emergence of so-called Lifshitz tail in the spectral density of one-dimensional random system with the Poisson disorder.

The paper is structured as follows. In Section II we formulate the model of an inflated planar ideal polymer loop and derive the corresponding partition function. We analyze the fluctuations of inflated loop in absence of a boundary and find that fluctuations are Gaussian. In Section III we suppress the large-scale fluctuations of the loop by inserting the impenetrable disc inside the inflated polymer and compute again the corresponding distribution function of chain monomers. We show the emergence of the KPZ-like scaling with the critical exponent $\gamma=\frac{1}{3}$ for relative fluctuations of monomers with respect to the boundary. The change of the exponent $\gamma$ with shrinking the radius of the inserted disc, is analyzed in Section IV. The discussion of obtained results and possible generalizations of the model are presented in Section V (Discussion).

\section{Partition function of inflated planar ideal polymer loop}

Consider the grand canonical ensemble of Brownian loops enclosing an algebraic area $S$ in the $(xy)$ plane, controlled by the constant magnetic field ${\bf B}=(0,0,2)$ perpendicular to the $(xy)$-plane. The corresponding Hamiltonian $H$ can be written as
\be
H = \frac{1}{a^2} \left(\nabla + iq {\bf A}\right)^2
\label{2}
\ee
where the vector potential ${\bf A}=\frac{1}{2}{\bf B}\times {\bf r}$ in polar coordinates $(r,\phi)$ has components $A_\phi=r$ and $A_r=0$. Evidently, $H$, is the Hamiltonian of a charged particle of mass $m=\frac{a^2}{2}$ moving in the plane $(xy)$ in the transversal magnetic field, ${\bf B}$.

In a homogeneous magnetic field, charged particles move along circular orbits with a certain angular (Larmor) frequency, $\omega$, where $\omega$ is defined by the well-known formula
\be
\omega=\frac{B|q|}{m}
\label{3}
\ee
In the framing, where $a=1$, $m=\frac{1}{2}$ and $B=2$, the charge, $q$, equals to
\be
q=\frac{\pi}{2T}=\frac{\pi}{2N}=\frac{c \pi}{2 R_g}\sim \frac{1}{R_g},
\label{4}
\ee
In \eq{4} we have supposed that the gyration radius, $R_g$, scales linearly with the chain length, i.e. $R_g = cN$ ($c<\frac{1}{2\pi}$) and the time of a flight around a disc, $T$, is equal to the number of steps, $N$, of a polymer, i.e. $T=N$.

The master equation, $\frac{\partial}{\partial t}P=HP$, for the probability distribution $P(r,\phi,t)$ of a charged particle in a constant magnetic field, has the following explicit form:
\be
\frac{\partial P(r,\phi,t)}{\partial t} = \left(\nabla^2 + iq(\nabla {\bf A}+ {\bf A} \nabla) - q^2 B^2\right) P(r,\phi,t)
\label{5}
\ee
Taking into account that $\left(\nabla {\bf A}+ {\bf A} \nabla\right) = 2\frac{\partial}{\partial \phi}$, we can rewrite \eq{5} as
\be
\frac{\partial P(r,\phi,t)}{\partial t} = \left[\frac{1}{r}\frac{\partial}{\partial r} \left(r\frac{\partial}{\partial r}\right) + \frac{1}{r^2} \frac{\partial^2}{\partial \phi^2} \right] P(r,\phi,t) - iq \frac{\partial}{\partial \phi} P(r,\phi,t) - q^2 r^2 P(r,\phi,t)
\label{6}
\ee
Let us suppose that all paths start at the point $A(r_0=R_g,\phi=0)$ and terminate at the point $B(r_1=R_g,\phi=2\pi)$, i.e. trajectories making a full turn around the center of coordinates, end at the same point $A$. Thus, the probability distribution, $P(r,\phi,t)$, is angularly-symmetric and, hence, is independent on $\phi$. Suppose also that there is an auxiliary small disc of a radius $R_d$ (with $R_d\to 0$) placed at the center of coordinates. We consider such a disc as a point obstacle. Defining $P(r,\phi,t) = Q(r,t)$, we get for $Q(r,t)$ the following boundary problem
\be
\begin{cases}
\disp \frac{\partial}{\partial t}Q_m(r,t) =  \left(\frac{\partial^2}{\partial r^2} + \frac{1}{r} \frac{\partial}{\partial r}  - q^2 r^2\right) Q_m(r,t) \medskip \\
\disp Q_m(r=R_d\rightarrow 0,t) = Q_m(r=\infty,t)=0 \medskip \\
\disp Q_m(r,t=0) = \frac{\delta_{r,R_g}}{r}
\end{cases}
\label{7}
\ee
Separating variables $r$ and $t$, write $Q_m(r,t)$ in the form $Q_m(r,t) = W_m(r)\, e^{-\lambda t}$, where $\lambda\ge 0$. The function $W(r)$ satisfies the set of equations
\be
\begin{cases}
\disp -\lambda W_m(r) = \left(\frac{\partial^2}{\partial r^2} + \frac{1}{r} \frac{\partial}{\partial r} - q^2 r^2\right) W_m(r,t) \medskip \\
\disp W_m(r=R_d\rightarrow 0,t) = W_m(r\rightarrow\infty,t)=0 \medskip \\
\disp W_m(r,t=0) = \frac{\delta_{r,R_g}}{r}
\end{cases}
\label{8}
\ee
The solution to \eq{8} vanishing at $r\rightarrow\infty$ is
\be
W_n(r) = e^{-\frac{q\, r^2}{2}}L_{n}(q r^2)
\label{9}
\ee
where $n=\frac{\lambda-2q}{4q}$ and $L_n(x)$ is the Laguerre polynomial. Now we can write the solution to \eq{7} with the $\delta$-initial condition in the following form:
\be
Q(r,t)=\sum\limits_{n=0}^\infty\frac{n!}{\Gamma(n+1)}e^{-4qt(n+1/2)} e^{-q(r-R_g)^2/2}L_n(q r^2)L_n(q R_g^2)
\label{10}
\ee
Using the properties of sums involving Laguerre polynomials, we get
\be
Q(r,t)=\frac{1}{1-e^{-4qt}}\exp \left(-\frac{q e^{-4qt} \left(r^2+R_g^2\right)} {1-e^{-4qt}}-\frac{1}{2} q \left(r-R_g\right)^2-2qt \right) I_0\left(\frac{2 q r R_g e^{-4qt}} {1-e^{-4qt}}\right),
\label{11}
\ee
It can be easily seen that in the limit of a large argument of a Bessel function, Eq. \eq{11} gets transformed to the following expression
\be
Q(r,t)\sim \frac{1}{1-e^{-4qt}}\exp \left(-\frac{q e^{-4qt} \left(r-R_g\right)^2} {1-e^{-4qt}}-\frac{1}{2} q \left(r-R_g\right)^2-2qt \right)
\label{12}
\ee

Returning to the initial Brownian bridge problem, we define the conditional probability, $Z$, for a Brownian bridge starting at the point $A$ to be at some intermediate point $(r,\phi)$ after time $t$, and to reach the point $B$ at time $T=2t$. Combining the expression for $Z$ with the condition \eq{4}, we get:
\be
Z(r,R_g) = {\cal N}^{-1}\ Q(r,t)^2={\cal N}^{-1} \exp \left(-\frac{c \pi \left(r-R_g\right)^2} {2 R_g}\frac{1+e^{-4\pi}}{1-e^{-4\pi}}\right);
\label{12a}
\ee
where ${\cal N}$ is the normalization constant
\begin{multline}
{\cal N} =  \int_0^{\infty} rdr\exp \left(-\frac{c \pi \left(r-R_g\right)^2} {2 R_g}\frac{1+e^{-4\pi}}{1-e^{-4\pi}}\right)=\\
=\frac{1}{2} \sqrt{\frac{e^{4 \pi }-1}{2 \pi c(1 + e^{4 \pi } ) }} R_g^{3/2} \left(\Gamma \left(-\frac{1}{2},\frac{c\pi  R_g \left(1+e^{4 \pi }\right) }{2 \left(-1+e^{4 \pi }\right)}\right)+4 \sqrt{\pi }\right)
\label{12b}
\end{multline}
where $\Gamma(a,x)=\int_x^{\infty}t^{a-1}e^{-t}dt$ is the incomplete gamma-function.

The variance of fluctuations, $\mathrm{Var}[r(R_g)]$, of inflated paths ($N=\frac{1}{c} R_g$), described by the Gaussian distribution function \eq{12a} at typical value $q=\frac{\pi}{2 N}$ (see \eq{4}) is:
\be
\sqrt{\mathrm{Var}[r(R_g)]}=\left[\frac{1}{{\cal N}}\int_{0}^{\infty} r^2\, Z(r,R_g) rdr -\left(\frac{1}{{\cal N}}\int_{0}^{\infty} r\,Z(r,R_g)r dr\right)^2\right]^{1/2}\propto R_g^{1/2}
\label{13}
\ee
where ${\cal N} = \int_0^{\infty} Z\left(r, R_g\right) rdr$ is the normalization of the distribution function.

\begin{figure}[ht]
\epsfig{file=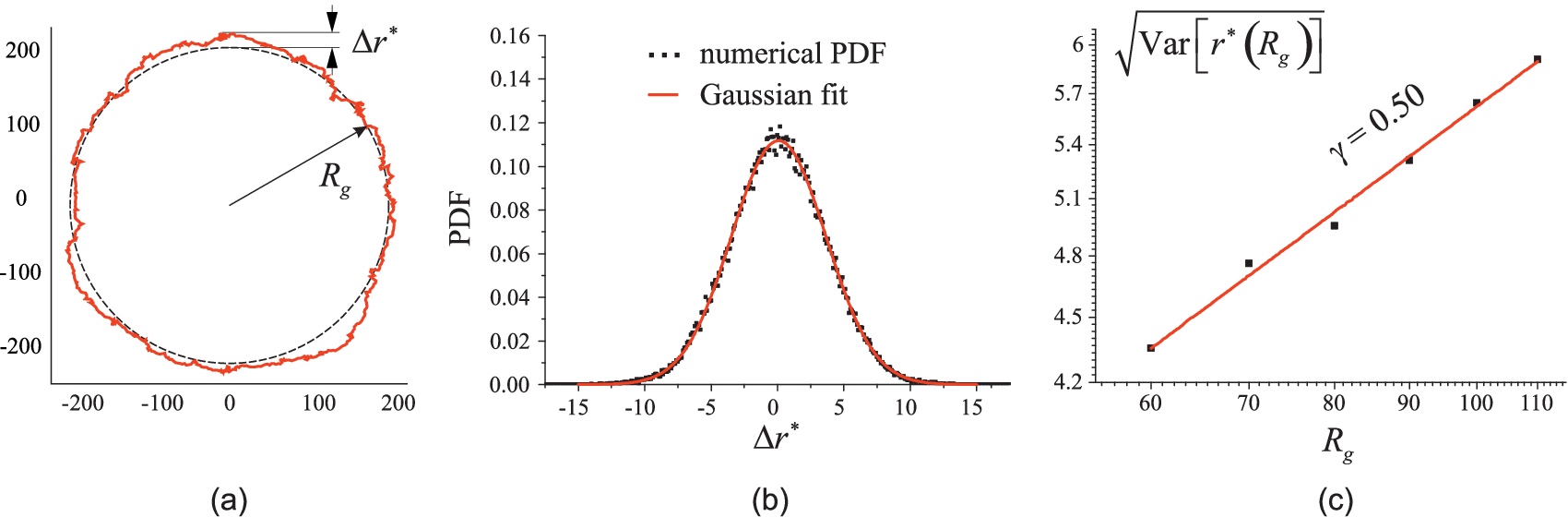, width=16cm}
\caption{(a) Strongly inflated ideal random loop fluctuating around the ``optimal'' circle of radius $R$; (b) Distribution of the fluctuations $\Delta r^*$ of the inflated ring with respect to the equilibrium shape and its comparison with $\mathcal{N}(0,\sigma^2)$, where $\sigma^2\approx 12.725$; (c) Gaussian scaling of fluctuations as a function of $R_g$ in log-log coordinates.}
\label{fig:01}
\end{figure}

The sample of an $N$-step strongly inflated Brownian loop with a large gyration radius, $R_g = \frac{N}{7}$, obtained in the numeric simulations, is shown in \fig{fig:01}a. The path is fluctuating around the ``optimal'' equilibrium shape which is a circle of radius $R_g$ depicted by the dotted line in \fig{fig:01}a. We are interested in typical radial deviations, $\Delta r$, of monomers of the random loop from the equilibrium shape (the dotted line) as a function of $N$ (or of $R_g$ since $R_g = \frac{N}{7}$).

From \eq{12}--\eq{13} one concludes that despite the trajectories are pushed by the ``loop inflation'' (i.e. by the path stretching) to an improbable tiny region of the phase space, the very presence of a large deviation regime is not sufficient to affect the path's statistics. In \fig{fig:01}b,c we have shown the results of the numeric simulations for the distribution function of monomers of inflated loop and the square root of the variance of their fluctuations (compare to \eq{13}). One clearly sees that fluctuations are Gaussian with the critical exponent $\gamma=\frac{1}{2}$.

\section{Statistics of inflated planar ideal polymer loop leaning on an impenetrable disc}

In order to suppress the large-scale fluctuations of a strongly inflated ideal polymer chain, we insert inside the polymer loop the impenetrable disc such that the Brownian path is ``leaning'' on the external disc boundary. The analytical solution mostly repeats the steps given in the previous section with one exception: the radius of inserted disc tends to the radius of gyration of the polymer, $R_d\rightarrow R_g$ .

Let us return to Eq. \eq{8} and make use of the substitution $W(r) = U(r)r^{-1/2}$. This permits us to rewrite \eq{8} in the following form:
\be
\begin{cases}
\disp \frac{d^2U(r)}{dr^2}- \left(q^2r^2-\frac{1}{4r^2}-\lambda \right) U(r) = 0 \medskip \\
\disp U(r=R_g,t) = U(r=\infty,t)=0
\end{cases}
\label{14}
\ee
We analyze the solution of \eq{14} in the vicinity of the disc boundary, i.e. when $r$ can be written as $r = R_g+\rho$, where $0<\frac{\rho}{R_g}\ll 1$. We get in this limit
\be
U''(\rho)- c_1 U(\rho) -c_2 \rho\, U(\rho) = 0
\label{15}
\ee
where
\be
c_1=q^2R_g^2-\frac{1}{4R_g^2}-\lambda, \qquad c_2=2q^2R_g+\frac{1}{2R_g^3}
\label{16}
\ee
Vanishing at infinity solution to \eq{15}  can be written in the  form
\be
U(\rho)\propto \mathrm{Ai}\left(\frac{c_2 \rho+c_1}{c_2^{2/3}}\right)
\label{15a}
\ee
where $\mathrm{Ai}(x)=\frac{1}{\pi}\int\limits_0^\infty \cos\left(\frac{t^3}{3}+x t\right)dt$ is the Airy function.

Combining the limit of large $R_g$ with the condition $q\sim\frac{1}{R_g}$ (see \eq{4}), we can rewrite \eqref{16} as follows:
\be
c_1=\frac{2^{2/3} a_i}{R_g^{2/3}}, \qquad c_2=\frac{2}{R_g} \qquad \lambda=1-\frac{2^{2/3} a_i}{R_g^{2/3}},
\label{16a}
\ee
so the solution to \eq{15}--\eq{16a} that satisfies zero boundary conditions is:
\be
U(\rho)\propto  \mathrm{Ai}\left(\left(\frac{2}{R_g}\right)^{1/3}\rho+a_i\right)
\label{17}
\ee
where $a_{i}<0$ are zeros of the Airy function, i.e. the solutions of $\mathrm{Ai}(a_{i})=0$ ($i=1,2,...$).

The general solution to the initial problem \eq{7} in the presence of the impenetrable disc of radius $R_d\rightarrow R_g$ reads:
\be
Q(r,t)=\sum\limits_{i=0}^\infty \frac{\mathrm{A_i}}{\sqrt{R_g+\rho}} e^{-\left(1-\frac{2^{2/3} a_i}{R^{2/3}}\right)t} \mathrm{Ai}\left(\left(\frac{2}{R_g}\right)^{1/3}\rho+a_i\right); \quad (i=1,2,...)
\label{17a}
\ee
where $A_i$ is chosen to fulfil the initial conditions.

We are interested in the behavior of the midpoint of the Brownian bridge, so the total travelling time, $t=T$ is of the order of the time of one complete turn around a disc, $T\sim N$. Since the series \eq{17a} decreases exponentially with a speed $\exp\left(-2^{2/3} |a_i| R_g^{1/3}\right)$ (see \eq{17a}) and $0<\frac{\rho}{R_g}\ll 1$, we can leave only the first term in \eq{17a}:
\be
Q(r,t)\approx (R_g+\rho)^{-1/2}e^{-\left(1-\frac{2^{2/3} a_1}{R^{2/3}}\right)t} \mathrm{Ai}\left(\left(\frac{2}{R_g}\right)^{1/3}\rho+a_1\right);
\label{17b}
\ee

For the initial Brownian bridge problem (see \eq{12a}), we can define the conditional probability, $Z$, for a midpoint of the Brownian bridge similarly to \eq{12a}:
\be
Z(r) = {\cal N}^{-1}\; r^{-1} \mathrm{Ai}^2 \left(\left(\frac{2}{R_g}\right)^{1/3}(r-R_g)+a_1\right)\approx{\cal N}^{-1}\;R^{-1}\mathrm{Ai}^2 \left(\left(\frac{2}{R_g}\right)^{1/3}(r-R_g)+a_1\right)
\label{18}
\ee
as $0<\frac{r-R_g}{R_g}\ll 1$  and ${\cal N}$ is the normalization constant
\be
{\cal N} =  \int_R^{\infty} dr\; \mathrm{Ai}^2 \left(\left(\frac{2}{R_g}\right)^{1/3} (r-R_g)+a_1\right)= \left(\frac{R_g}{2}\right)^{1/3}\mathrm{Ai}'(a_1)^2
\label{19}
\ee

Eqs. \eq{18}--\eq{19} provide the explicit expression for the variance of fluctuations of the inflated loop  (for which $N=\frac{1}{c} R_g $) above the impenetrable disc of radius $R_d=R_g$:
\be
\sqrt{\mathrm{Var}[r(R_g)]}\sim R_g^{1/3}
\label{21}
\ee

\begin{figure}[ht]
\epsfig{file=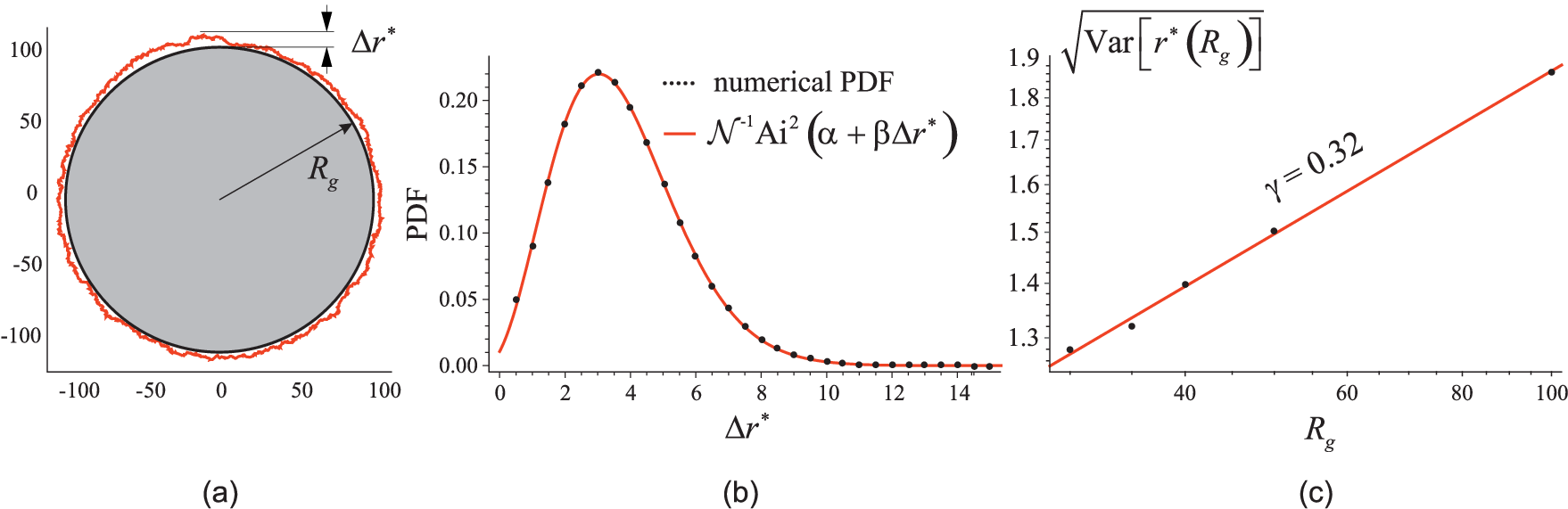, width=16cm}
\caption{(a) The impenetrable disc of the radius $R_d=R$ is inserted inside the strongly inflated ideal random loop of gyration radius $R$ such that the path is leaning on a disc boundary; (b) Distribution of the fluctuations of the ring above the impenetrable disc fitted by the square of the Airy function ${\rm Ai}^2(\alpha+\beta\Delta r^*)$, where $\alpha\approx -2.195$ and $\beta \approx 0.239$; (c) KPZ scaling of fluctuations as a function of the disc radius, $R_d$, in log-log coordinates.}
\label{fig:02}
\end{figure}

Figure \ref{fig:02}a presents the snapshot of computer simulations of a particular realization of a single stretched random path evading the solid impenetrable disc of radius $R_d=R_g$. Computing the typical span, $\Delta r^*$, of fluctuations of monomers above the disc, we see from \eq{18} that fluctuations are essentially non-Gaussian and are controlled by the Kardar-Parisi-Zhang critical exponent $\gamma=\frac{1}{3}$. This conclusion is fully consistent with numeric simulations shown in \fig{fig:02}b,c.

According to \eq{18}--\eq{21}, the typical span, $\Delta r^*$, of trajectories avoiding circular domain possess the Kardar-Parisi-Zhang scaling: $\Delta r^* \propto N^{1/3}$, i.e. paths get confined near the disc boundary within a circular strip of typical width $\propto N^{1/3}$. In \fig{fig:02}b we have plotted the distribution function of $\Delta r^*$, which perfectly coincides with the square of the Airy function in \eq{18}. In \fig{fig:02}c we have depicted the variance of fluctuations of inflated Brownian loop leaning on an impenetrable disc, $\mathrm{Var}^{1/2}(R_g) \propto R_g^{\gamma}$ with $\gamma \approx 0.32$. This value of $\gamma$ is very close to the KPZ critical exponent, $\gamma=\frac{1}{3}$ and is in full agreement with \eq{21}. Let us remind that in absence of the hard-wall constraint, the fluctuations of the inflated loop are Gaussian which we do see in numeric simulations shown in \fig{fig:01}b,c.

We argue that simultaneous fulfilment of two conditions which restrict the large-scale fluctuations of an ideal polymer chain: (i) the path stretching (i.e. the loop "inflation"), and (ii) the hard-wall convex constraint (i.e. insertion of an impenetrable disc), are crucial for the localization of path's fluctuations within the strip of width $N^{1/3}$ near the disc boundary. By stretching, trajectories are pushed to an improbable tiny region of the phase space. However, as we have seen in Section II, the very fact of a large deviation regime is not sufficient to affect the path's statistics and the presence of a solid convex boundary on which paths are leaning, is necessary. The importance of a convex boundary has been emphasized in \cite{peres,shlosman}.

\section{2D random walk winding around a disc}

The problem of angular wandering of a 2D random walk was the subject of many works (see, for example \cite{grosberg} for a review). The authors were mainly interested in a winding angle distribution of a planar polymer chain in presence of an impenetrable obstacle or a finite disc. Recently, in \cite{vladimirov} the stationary distribution of a random walk radial density has been studied in a similar system, however under the condition that a walk has an angular drift. It has been shown in \cite{vladimirov} that the stationary distribution is given by squared Airy function with the typical KPZ-type scaling. Our results are consistent with the results of \cite{vladimirov}, as well as with the ones obtained for stretched random walks above the circular voids in one-dimensional \cite{f-s} and two-dimensional \cite{polov} geometries. Also, our consideration rhymes with the results of the work \cite{peres}, were authors studied fluctuations of Brownian loops covering an atypically large area.

Due to the strong impact of entropic effects, free Brownian loops typically belong to the Gaussian class of universality with $\gamma=\frac{1}{2}$. As we have seen in Section II, imposing a constraint on the area covered by the 2D random walk, which force the random loop to stay in the large deviation regime, does not produce any influence on scaling behavior of fluctuations. Even such topological constraints as formation of local knots on a random path \cite{MetzlerKardar}, do not bring the system out of the Gaussian universality class.

\begin{figure}[ht]
\epsfig{file=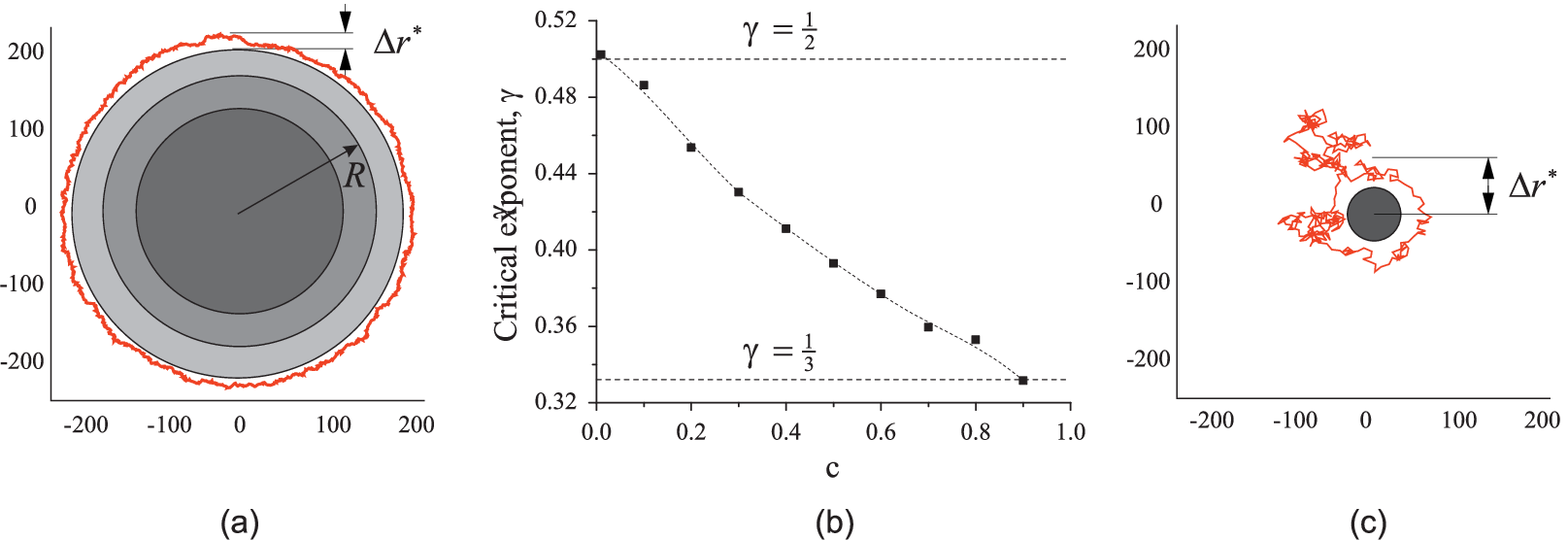, width=16cm}
\caption{(a) Polymer loop of length $L=7R_g a$ ($a=1$) leaning on an impenetrable disc of changing radius $R=cR_g$; (b) Dependence of the critical exponent $\gamma$ on $c$, where $\gamma(c)$ is defined in \eq{Delta}; (c) Typical snapshot of the system for $c\approx 0.1$.}
\label{fig:03}
\end{figure}

Meanwhile, entropy can compete with geometric constraints forcing the random walk to follow atypical paths close to trajectories emerging in the ``geometric optic'' approach \cite{meerson1, meerson2}. To demonstrate this, we consider a random walk of $N=7R_g/a$ steps (where $a=1$ for simplicity) conditioned that the path is ``leaning'' on an impenetrable disc of radius $R=cR_g$ as it is schematically shown in \fig{fig:03}a with $c$ changing from $\approx 1$ (almost fully inflated ring) down to 0 (the point-like obstacle). We investigate the dependence of the scaling exponent $\gamma(c)$ on $c$. Recall that $\gamma(c)$ is defined by the relation
\be
\Delta r^*(N) \propto N^{\gamma(c)}
\label{Delta}
\ee
 Note, that by stretching condition, $R = c R_g = \frac{c}{7} N$. Thus, \eq{Delta} is equivalent to scaling relation $\Delta r^*(R) \propto R^{\gamma(c)}$. The corresponding plot $\gamma(c)$ for $c$ changing from 1 down to 0 is shown in \fig{fig:03}b. The typical plot of the system for $c \approx 0.05$ is depicted in \fig{fig:03}c.

In the limit $c\searrow 0$ the disc is shrinking to a point, and the entropy gradually suppresses the effect of a boundary constraint. Fluctuations return to the Gaussian regime:
\be
(\Delta r^*)^2 =\frac{a^2 N}{6}\left(1+\frac{3}{2\pi n^2} \right)
\label{22}
\ee
where $n$ is the winding number around the point (in our case $n=1$) and $a$ is the monomer length (see, for example, \cite{grosberg} for more details).

In the opposite case, $c\nearrow 1$, discussed at length of Section III, the disc occupies almost all conformational space, forcing the strongly inflated loop to stay in a tiny region of a phase space near the disc boundary, where the role of entropy is essentially suppressed by the geometric constraint. Note, that in absence of a disc, the spontaneous realization of such fluctuations is very improbable. So, the random walk statistics at $c\approx 1$ is controlled both by the system geometry and the entropy, which is manifested by the KPZ-like scaling exponent, $\gamma=\frac{1}{3}$ for $\Delta r^*$ in \eq{Delta}. The "longitudinal" fluctuations (along the disc boundary) are controlled by the critical exponent $\gamma_{\parallel}=2\gamma=\frac{2}{3}$.

At intermediate values of $c$ there is a competition between the entropy and the geometry (recall that still the length of the loop is $Na = 7R$: at small $c$ there are $\sim (N-cR)\gg cR$ ``free'' monomers that do not participate in encircling the disc and can freely fluctuate, while at large $c\approx 1$ only $\sim (N-cR)\ll cR$ ``free'' monomers fluctuate competing with the curved geometry of the disc boundary.

\section{Conclusion}

We have shown that ensemble of ``inflated'' two-dimensional Brownian loops with a gyration radius, $R_g$, which is equal to the radius $R_d$ of an impenetrable disc inserted inside the loop (i.e. at $R_g=R_d$), demonstrates non-Gaussian fluctuations belonging to the Kardar-Parisi-Zhang (KPZ) universality class with the critical exponent $\gamma = \frac{1}{3}$. To the contrary, if one fixes the same degree of the loop inflation controlled by $R_g$, however remove the inserted hard disc by setting $R_d\to 0$, the fluctuations return to the Gaussian regime with $\gamma = \frac{1}{2}$.

The physical origin of drastic change of $\gamma$ deals with the presence/absence of long-wave fluctuational modes: suppressing soft long-range spatial fluctuations by a hard-wall constraint (such as an impenetrable disc) for inflated polymers, we push the system out of a Gaussian regime towards the regime controlled by KPZ fluctuations. The setup of the system which has been treated both analytically and by direct Monte-Carlo simulations, is rather simple: we take a two-dimensional random loop of length $L=Na$, fix its gyration radius, $R_g$, such that $R_g \lesssim \frac{Na}{2\pi}$ and insert inside the inflated loop the impenetrable disc of changing radius $R_d$.

It should be pointed out that KPZ fluctuations are not universal and depend on the geometry of the object encircled by the Brownian loop. For example, if one inserts inside the inflated random loop the convex figure whose boundary is not circular, but for example, is determined by some algebraic curve of higher order, the fluctuations will have different critical exponents above different points of the boundary -- see \cite{polov} for more details.

As concerns further developments of the model, it would be interesting to study in more details intermediate regimes when inserted hard disc of radius $R_d$ is of moderate size compared to $R_g$. Also, the investigation of statistics of inflated non-selfavoiding loops (and in general, of polymers with any fractal dimentsion $\nu$) seems to be a very challenging problem. Whether KPZ scaling survives for fluctuations of inflated fractal polymers running above the hard disc is an question which is addressed in a work in progress \cite{grosberg2}.

\begin{acknowledgments}

We are grateful to A. Gorsky, N. Mazotov, K. Polovnikov, S. Shlosman, M. Tamm, A. Vladimirov and V. Avetisov for numerous discussions and valuable comments. Numerical simulations shown in \fig{fig:01} and \fig{fig:02} and derivation of \eq{12a}--\eq{13} have been carried out by AV and are supported by the BASIS Foundation in frameworks of the grant 19-1-1-48-1. The derivation of \eq{18}--\eq{21} and the analysis of the interplay of large deviation regime and geometric constraints have been performed by SN in the framework of the RSF grant 21-11-00215.

\end{acknowledgments}

\end{document}